# EXPLORATION OF THE OUTER SOLAR SYSTEM WITH FAST AND SMALL SAILCRAFT

National Academy of Sciences
The Planetary Science and Astrobiology Decadal Survey 2023-2032

## A WHITE PAPER

Slava G. Turyshev[1], Peter Klupar[2], Abraham Loeb[3], Zachary Manchester[4],
Kevin Parkin[2], Edward Witten[5], S. Pete Worden[2]

[1]*Jet Propulsion Laboratory, California Institute of Technology,
4800 Oak Grove Drive, Pasadena, CA 91109-0899, USA*

[2]*Breakthrough Initiatives, Building 18, Second Floor, P.O. Box 1
Moffett Field, CA 94035-0001, USA*

[3]*Astronomy Department, Harvard University, 60 Garden Street, Cambridge, MA, USA*

[4]*Stanford University, Stanford CA 94305, USA*

[5]*Institute for Advanced Study, Einstein Drive, Princeton, NJ 08540 USA*

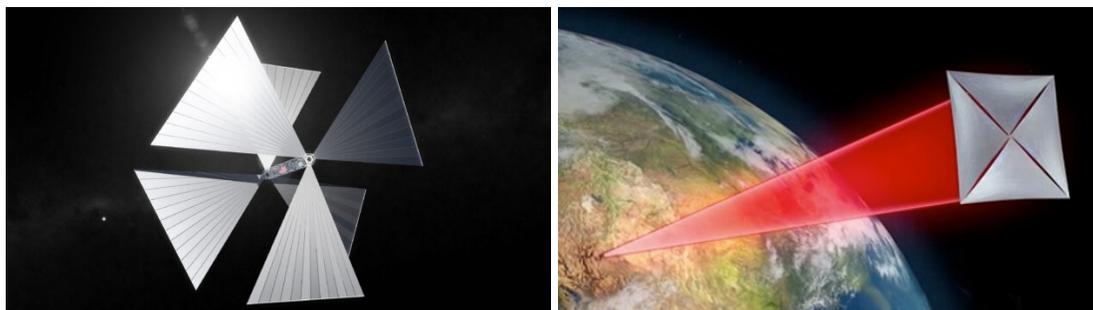

Left: An artist's impression of a solar sail for the Solar Gravitational Lens Focus mission[1].
Right: An artist's impression of a Breakthrough Starshot spacecraft in action.

---

[1] https://www.xplore.com/press/releases/2020/04.28.2020_Xplore_SGLF_NASA_NIAC_Phase_III.html

## 1. New Decade in Planetary Science: Exploring Solar System Faster and Farther

Chemical propulsion limits the distance that we can reach to ~50AU (Pluto's distance) within a mission travel time of a decade, or ~500AU (Planet 9's distance) within a century. Achieving higher speeds through alternative launch methods could substantially reduce these durations.

Two new interplanetary technologies have advanced in the past decade to the point where they may enable exciting, affordable missions that reach further and faster deep into the outer regions of our solar system: (i) Interplanetary smallsats, developed and flown by JPL as MarCO[2] on the Mars InSight mission; and (ii) Solar sails, which utilize solar radiation pressure for propulsion. The successful JAXA-built spacecraft IKAROS[3] demonstrated solar sail technology on a Venus-bound mission launched in 2010. In 2019, successful orbital demonstration by the LightSail-2[4] flight led by The Planetary Society raised confidence in solar sails and paved the way for two of NASA interplanetary missions, NEA-Scout[5] and Solar Cruiser[6], which are planned for the near future. Japan is now developing OKEANOS[7] as a follow-up to IKAROS for outer planet missions.

A related concept is the laser sail, in which a higher intensity photon beam is generated on Earth. One such mission concept is the *Breakthrough Starshot Initiative*[8] led since 2016 by S. Pete Worden and Avi Loeb (who are co-authors on this proposal) and supported by Yuri Milner, Mark Zuckerberg and the late Stephen Hawking. This effort strives to lay the groundwork for a mission to the Alpha Centauri star system within the next generation. The concept builds on two major ideas: shrink present-day spacecraft to a total mass of ~1 g and leave the fuel on the ground. With such a system, spacecraft velocities of up to 20% of the speed of light are potentially feasible.

With sailcraft velocities approaching a fraction of the speed of light, travel times within the solar system are dramatically reduced (Lubin & Hettel, 2020). Once the development and construction of the necessary ground-based infrastructure is accomplished, launch costs are reduced to the maintenance cost of the Earth-based infrastructure and the energy required to accelerate individual spacecraft. This may open up deep space for a much more diverse scientific audience, similar to what CubeSats have accomplished during the last decade (Millan et al., 2019).

An Earth-based laser propulsion system consists of a laser array and is thus fully scalable. Ramping up the accelerative force only requires adding extra lasers to the photon engine, thus offering capabilities for solar system exploration even before the system is fully operational.

## 2. Fast access to outer solar system with sailcraft

Small and capable sailcraft could be placed on fast solar system transit velocities. Cruise velocity of 300 km/s ($10^{-3}c$) enables a sailcraft to reach Pluto within a year, the heliosphere in 2 years, and interstellar space in 5 years. This velocity can be achieved by a laser-pushed sailcraft mission at a more modest cost than the cost of a $0.2c$ Starshot mission (Parkin 2018). These costs are further expected to decrease as a result of improved technologies, mass production and facility reuse.

If a sailcraft is accelerated out of the ecliptic plane in the direction of the Earth's North or South poles, and if occasional cloud cover can be tolerated, then a single terrestrial facility can be used for the entire time (see Appendix). Otherwise, to maintain a beam for more than a few hours, there need to be several facilities that hand off from one to the next as the Earth rotates or there is cloud cover. Alternatively, a single beam facility may accelerate the sailcraft for a fraction of each day.

---

[2] https://www.jpl.nasa.gov/cubesat/missions/marco.php
[3] https://global.jaxa.jp/countdown/f17/overview/ikaros_e.html
[4] https://www.planetary.org/explore/projects/lightsail-solar-sailing/
[5] https://www.nasa.gov/content/nea-scout/
[6] https://en.wikipedia.org/wiki/Solar_Cruiser
[7] https://en.wikipedia.org/wiki/OKEANOS
[8] https://breakthroughinitiatives.org/initiative/3

During the earlier phases the project, some elements of sailcraft operations and control may be tested with smallsats propelled by solar sails. With solar sails, a trajectory may start from Earth orbit, spiraling in toward the Sun and then, after passing through solar perihelion, transition to a high energy hyperbolic trajectory to the outer solar system and beyond. With today's sail technology we can achieve speeds of ~7 AU/yr (35 km/s) (Turyshev et al., 2020). Demonstration of these capabilities is one of the primary objectives of the ongoing NIAC Phase III effort (Xplore, 2020).

Much higher solar sailing velocities, potentially up to 300 km/s, may be achieved with a shorter perihelion distance, where the sailcraft must withstand the high radiation flux emanating from the Sun. The minimum heliocentric distance at perihelion will be determined by the sail material. Materials including ceramic dielectrics, such as silicon nitride and silica possess high melting temperatures $\geq 2{,}000$K, while having low mass density and low solar absorptivity. Sails made of ceramic dielectrics may reach a perihelion of ~2−5 solar radii without heating above their melting point, thus providing a pathway for planetary missions with higher solar system exit velocities.

Such a phased approach opens a new paradigm to develop missions through the solar system that enable exploration of distant regions of the solar system to occur decades sooner and faster than previously considered. The paradigm involves potentially multiple smallsats and ultrafast initially solar- and then laser-propelled trajectories that create a series of increasingly ambitious missions moving us closer to the stars. Prior to interstellar missions, the testing of the sailcraft system and design within the solar system will allow for a new branch of scientific studies.

Yet another approach is to tap into the momentum carried by the solar wind by using a mesh of wires that is offered by the electric sail approach. Such a system can reach $0.001c$, comparable in magnitude to the speed of the solar wind. The scaling laws associated with the electric sail concept were analyzed by Lingam & Loeb (2019).

## 3. Science Missions Enabled by Fast Transit Through the Solar System

### 3.1 *Chasing After and Probing Interstellar Objects*

In October 2017, the first known interstellar object (ISO) to visit our solar system was discovered. The object, named 1I/'Oumuamua, was detected, tracked, and observed within 2 weeks as it was moving through the solar system at a heliocentric velocity of ~50 km/s. This discovery allowed for a calibration of the abundance of interstellar objects of its size and an estimation of the subset of objects trapped by the Jupiter-Sun system (Siraj & Loeb, 2019). There should be thousands of 'Oumuamua-size interstellar objects identifiable by Centaur-like orbits at high inclinations, assuming a number density of 'Oumuamua-size interstellar objects of $\sim 10^{15}$ pc$^{-3}$. In fact, there are eight known objects that may be of interstellar origin. In addition, in Aug 2019, 2I/Borisov, became the first observed interstellar comet and the second observed ISO after 'Oumuamua. 2I/Borisov has a heliocentric orbital eccentricity of 3.36 and is not bound to the Sun.

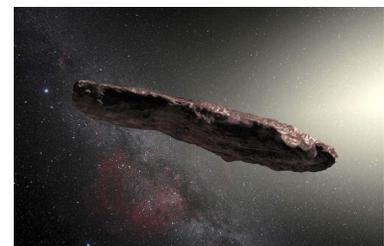
Artist's impression of 'Oumuamua.

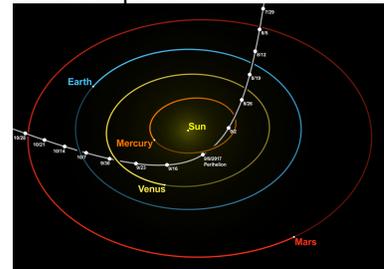
'Oumuamua's trajectory in the inner solar system.

It is likely that many of these objects pass undetected through the solar system every year. It is believed that there may be up $\sim 10^4$ of similar ISOs trapped inside the orbit of Neptune (Jewitt et al, 2017; de la Fuente Marcos et al., 2018). Photographing or visiting these objects and conducting *in situ* exploration would allow us to learn about the conditions in other planetary systems without sending interstellar probes.

Sailcraft on high-energy trajectories provide a unique opportunity to directly study ISOs transiting through the solar system. The scientific return from such investigations is invaluable, as comparative studies between an ISO sample return with solar system asteroid and comet sample returns can help us understand the conditions and processes of solar system formation and the nature of the interstellar matter—the first priority question listed in the Planetary Sciences Decadal Survey[9]. With many new ISOs expected, this topic should be of the highest priority for the new decade.

Rendezvous with an ISO will help to answer unsolved mysteries surrounding these interstellar interlopers. For example, to determine the origin of 'Oumuamua's non-gravitational acceleration: is it due to cometary outgassing, due to fractal dust aggregation formed over time, due to interstellar dust coalescing upon impact, or is it because of some other exotic origin? The interdisciplinary context of these investigations in a planetary science mission could lead to major progress in several areas of astrophysics. Fractal aggregates, which have been indirectly observed in circumstellar disks, may be the building blocks in protoplanetary disks thus yielding clues for processes guiding formation and evolution of planetary systems in our stellar neighborhood. Sending sailcraft to transient ISOs may allow us to directly access and study the building blocks of exoplanets, which could provide unprecedented constraints on planet formation models.

A close encounter by a swarm of sailcraft with an ISO would help us answer important questions such as: What caused these objects to accelerate? Is there outgassing and if so, what is causing it? What materials constitute these objects, and how do they compare to typical solar system asteroids or comets? Or perhaps these ISOs are fractal dust aggregates with such a low density that they can be accelerated by radiation pressure? These questions are fundamental, as the microscopic properties of dust played a key role in particle aggregation during the formation of our solar system.

With ISOs being distant messengers from interstellar space, we have an unprecedented opportunity to discover what these transient objects can tell us about our own solar system, its planetary formation and the interstellar medium in order to test theories regarding planetary formation. Such discoveries may lead us to reconsider comet and planetary formation models and inform about the properties of the transient interstellar object's parent nebula via its composition.

A very broad range of measurements are sought for ISOs. They include the characterization of basic physical properties (shape, density, morphology, dynamical properties), compositional properties (elemental composition, mineralogy, isotopes of at least hydrogen, oxygen, nitrogen, and carbon), geophysical/interior properties (porosity, cohesion, magnetic field), geological traits that might inform on origin and possible long-term evolution.

Small sun- or laser-propelled sailcraft are the only means of exploration we have that could catch up with ISOs moving at tens of km/s and be inexpensive enough to be on standby in Earth orbit. Although their speed may be far from the ultimate design speed of $0.2c$ (60,000 km/s) for Starshot, solar sailing or earlier demo versions of the laser-pushed sailcraft could represent a good intermediate step. If, for example, a 100 g spacecraft is accelerated to ~1/1000th of the ultimate speed, roughly 60 km/s, it could have made a rendezvous with 'Oumuamua at closest approach in ~8 days, assuming that 'Oumuamua had been spotted in time. If a pursuit had been initiated at the time when 'Oumuamua was actually discovered, it would have taken about a month to catch up with it. This capability is within reach today and should be considered for planetary exploration.

---

[9] https://www.nap.edu/catalog/13117/vision-and-voyages-for-planetary-science-in-the-decade-2013-2022

## 3.2 *Reaching the Solar Gravitational Lens for Direct Imaging of Exoplanets*

Imaging of extrasolar terrestrial planets combined with spectroscopy is probably the single greatest remote sensing result that we can contemplate in terms of galvanizing public interest and support for deep-space exploration. However, direct multipixel imaging of exoplanets requires significant light amplification and high angular resolution. With classical optical telescopes and interferometers, we face the sobering reality: i) to capture even a single-pixel image of an Earth-like exoplanet at a distance of 30 pc, a ~90 km telescope aperture is needed (for $\lambda = 1$ μm); ii) interferometers with thousands of telescopes (~30m) and baselines (~90 km) (Labeyrie 1996, 1999) will require integration times of ~$10^5$ years to achieve SNR=7 against the exozodiacal background. These scenarios are impractical, giving us no hope to spatially resolve and characterize exolife features.

To overcome these challenges, we may consider the Solar Gravitational Lens (SGL) as the means to produce high-resolution, multipixel images of exoplanets (Turyshev & Toth, 2017, 2020). This lensing results from the diffraction of light by the solar gravitational field, which focuses incident light at distances >548 AU behind the sun. The properties of the SGL are quite remarkable, including light amplification of ~$10^{11}$ and angular resolution of ~$10^{-10}$ arcsec.

A meter-class optical telescope with a modest coronagraph operating in SGL's focal region, beyond 650 AU, can yield a 200×200-pixel image of an Earth-like exoplanet at 30 pc in just 12 months, which is not possible by other known means. Even in the presence of the solar corona, the signal is high enough to image of such an object with ~50 km-scale resolution of its surface, enough to see signs of habitability, observe seasonal changes, image surface topography, and obtain spectroscopy of the atmosphere.

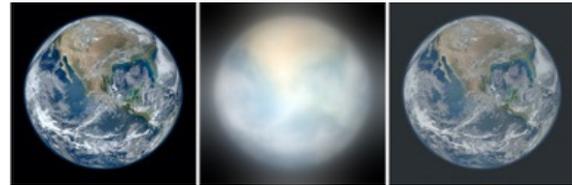

Imaging simulation with the SGL. Left: original RGB color image of an Earth-like exoplanet with a (1024x1024) pixels; center: image blurred by the SGL, sampled at an SNR of ~$10^3$ per color channel, or overall SNR of 3x$10^3$; right: image deconvolution result.

This makes the SGL is the only technique that could result in *direct* high-resolution, multipixel imaging of exoplanets. It is our only means to obtain resolved spectroscopic data that could show the presence of current life and to obtain an exo-Earth "close-up" images in the foreseeable future. Turyshev et al. (2020), have developed a new mission concept to deliver optical telescopes to the SGL's focal region and then to fly along the focal line to produce high resolution, multispectral images of a potentially habitable exoplanet.

The multismallsat architecture uses solar sailing and is designed to perform observations of multiple planets in a target extrasolar planetary system. It allows to reduce integration time, to account for target's temporal variability, which helps to "remove the cloud cover".

The prospect of getting an image of an exoplanet and to spectroscopically detect and characterize life there is compelling. With the SGL, we can search for possible biosignatures on planets where life similar to ours would have emerged and modified the atmosphere in a way that can be detected by remote sensing, i.e., spectroscopic observation of $O_2$, $O_3$, $H_2O$, $CO_2$, and $CH_4$ gases. If a solar sailing mission to the SGL focal region can provide spectroscopic proof of life on an exoplanet, it would qualify as one of the most exciting advances of scientific discovery in history.

## 3.3 *Probing the Plumes of Europa and Enceladus*

Europa is one of the most compelling astrobiological targets in the solar system. The combination of geological, compositional, gravitational, and induced magnetic field measurements all indicate a contemporary global saline liquid water ocean of ~100 km in depth (Sparks et al., 2016). Re-cently, Huybrighs et al. (2020) investigated decrease of fast protons during Europa flyby by Galileo spacecraft. They found that there is a special decrease, which can be explained by an erupting plume of water vapor, thereby providing strong evidence for an active watery plume on Europa.

Similar evidence mounts for gigantic water plumes on Enceladus. In 2005, NASA's Cassini spacecraft detected plumes of water vapor and icy particles erupting from Enceladus, revealing the existence of a giant ocean hidden under the moon's frozen shell. Because there is life virtually wherever there is water on Earth, these findings suggest that life might also exist on Enceladus. In 2018, the Cassini team reported detection of complex organic molecules from the moon, including some at least 15 carbon atoms in size. Clearly, *in situ* investigations of the plumes are critical.

Sailcraft could probe the plumes on Europa and Enceladus for complex molecules that are indicative of life. Solar sailing smallsats could explore what the recent discoveries say about the chances of life within these icy moons of Jupiter and Saturn. To target the plumes, one could deploy many sailcraft (sterilized in advance to protect from planetary contamination) in multiple sites to inform on the presence of organic molecules in the plumes. The swarms of small sailing spacecraft could provide opportunities to significantly enhance infrequent interplanetary missions with, e.g., landers or sacrificial satellites, and networks of small satellites that could enable missions to these unique objects to detect and study life that may exist on these bodies in the outer solar system. Swarms of inexpensive sailcraft could perform initial pioneering observations of these watery worlds and inform the development of future systems dedicated for their exploration.

### 3.4  *Studying the Shape and Structure of the Heliosphere*

As the Sun travels through the interstellar medium (ISM) it ejects plasma with supersonic speeds of 400–800 km/s called the solar wind. The solar wind flows well beyond the orbits of the planets and collides with the ISM. The plasma bubble-like region created by the solar wind around the Sun is called the *heliosphere*. At the heliosphere boundary interaction of the solar wind with the interstellar gas creates an interface with a complex structure.

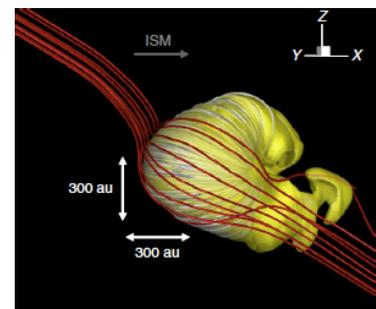

Two-lobe structure of the heliosphere. The white lines represent the solar magnetic field. The red lines represent the interstellar magnetic field (Opher et al., 2020).

The long-accepted view of the shape of the heliosphere is that it is a comet-like object with a long tail opposite to the direction in which the solar system moves through the local ISM (LISM). The solar magnetic field at a large distance from the Sun is azimuthal, forming a spiral as a result of the rotation of the Sun. The traditional picture of the heliosphere as a comet-like structure comes from the assumption that, even though the solar wind becomes subsonic at the termination shock as it flows down the tail, it is able to stretch the solar magnetic field. Based on magnetohydrodynamic (MHD) simulations, Opher, Loeb, Drake & Gabor (2020) established that the twisted magnetic field of the Sun confines the solar wind plasma and drives jets to the north and south very much like some astrophysical jets with the tension force being the primary driver of the outflow.

The shape of the heliosphere and the extent of its tail are thus subject to debate. According to the new model, as the Sun moves through the surrounding partially ionized medium, neutral hydrogen atoms penetrate the heliosphere, and through charge exchange with the supersonic solar wind, create a population of hot pick-up ions (PUIs). Until recently, the consensus was that the shape of the heliosphere is comet-like. The termination shock crossing by Voyager 2 demonstrated that the heliosheath (the region of shocked solar wind) pressure is dominated by PUIs; however, the impact of the PUIs on the global structure of the heliosphere has not been explored. The new model (Opher et al., 2020) reproduces both the properties of the PUIs, based on the New Horizons observations, and the solar wind ions, based on the Voyager 2 spacecraft observations as well as the solar-like magnetic field data outside the heliosphere at Voyagers 1 and 2.

It is therefore crucial to revisit LISM with modern, new *in situ* observations, which will be crucial to distinguish among the existing theories and to understand the physical picture of this region.

The main science questions that can be answered with sailcraft sent to various directions: How does the solar wind interact with the ISM and how does this relate to the interaction of other stars with their interstellar surroundings and formation of stellar astrospheres? How does this interaction lead to the observed complexities of the three-dimensional structure of the heliosphere? What is the nature of the termination shock? What is the nature and the processes that govern formation of the heliosheath? What are the properties of the heliopause transition region? How does the heliosphere affect the properties of very local ISM and how do they relate to the pristine ISM? Sailcraft could probe these questions related to the transition region from local to pristine ISM.

### 3.5 *Studies of the Pristine ISM*

The interstellar medium immediately outside our solar system is the closest example of cosmic *terra incognita*. As such, it should deserve a special place in our exploration agenda. However, the places that we explored directly with probes so far have essentially all been limited to the solar system and the interplanetary medium which is dominated by phenomena originating from our Sun (Stone et al., 2015). The gas and dust drifting among the stars, the ISM, is found throughout our Galaxy and is an integral part of energy flows and the material cycles thus defining the ecology of all galaxies. The ISM is the repository of the raw materials that are used to form new stars, which is replenished by stars at the end of their evolutionary cycles at which point they redeposit large quantities of material as supernovae or planetary nebulae.

The LISM, the material present in the immediate vicinity of the Sun, is the outer boundary condition that dictates the interaction of the Galaxy with the Sun. This interaction forms the heliosphere. Given the relatively large variability in ISM density and the relatively small variability in solar emission, the ISM dominates the general structure of the heliosphere – a boundary shielding us from the ISM and limiting flux of cosmic rays to the solar system. It is therefore of critical importance to fully characterize its properties. In particular, it is important to sample the ISM that has yet to be perturbed by the interaction with the Sun. This pristine ISM is located beyond the bow wave/shock, at a distance of >500 AU. The heliosphere acts as a filter, and there are components of the ISM that make it into the inner solar system. However, most of the material that makes up the ISM can only be sampled in its pristine form beyond the heliosphere.

In this vast volume of space, immediately beyond the heliosphere, where we find the pristine ISM, we also find the closest stars and planetary systems. Thus, even in our most immediate cosmic neighborhood, perhaps within a sphere of 10 pc, one finds a complex morphology of ISM clouds, hundreds of stars, dozens of known exoplanets, and a handful of astrospheres (structures analogous to the heliosphere around other stars). Accounting for the fact that stars with winds, orbited by planets, and adrift in the ISM are ubiquitous, the study of our own heliosphere and its interaction with the Galaxy will allow for understanding of all analogous structures.

Direct investigations of the LISM would provide critical information for understanding chemical evolution and mixing of matter within galaxies (e.g., neutrals, ions, isotopes, molecules, dust). Sampling each matter constituent would allow getting the fundamental elemental abundances, as current line-of-sight observations suffer from limited observational capability for some elements.

Therefore, *in situ* measurements are vital as they provide access to data that will have a critical impact on evaluating degeneracies or uncertainties in the long sight line average.

Subrelativistic sailcraft offer good opportunities to address these science questions. Getting to the pristine ISM is truly the next frontier in terms of unexplored territory. No other spacecraft have been in this region and making these kinds of *in situ* measurements is unique. Therefore, the scientific return is expected to be tremendous and transformative for ISM science.

### 3.6 *Studying Interstellar Turbulence*

As shown by Hoang & Loeb (2020), relativistic sailcraft offer unique opportunity to study the interstellar turbulence due to dust and particle content in the ISM, amplified by magnetic field.

Dust grains, which come in a wide range of sizes (from $10^{-3}$ to 1 micron), are ubiquitous components of the ISM and play an important role in interstellar chemistry, which results in the formation of the most abundant molecule ($H_2$), and a rich collection of complex organic molecules. Complete understanding of nearby dust is important to a broad range of areas of astrophysics, as it is a foreground contaminant of more distant signals. For instance, studying the aftermath of the Big Bang through the Cosmic Microwave Background (CMB) requires accurate removal of foreground dust. Critical measurements include composition, kinematics, and grain size distribution, with an emphasis on the complex organic molecules. Directly probing the volume density and small-scale structure are also critical to understanding the size and structure of the heliosphere.

Measurements to be conducted from a sailcraft would have an important astrobiological connection. They can be used to evaluate the formation and survivability of dust, and thereby illuminate the sites of chemical reactions that lead to the complex organic molecules that we find in the ISM.

The magnetic field threading through the LISM has a profound influence on the structure of our heliosphere and the distribution of many particles (e.g., energetic neutral atoms, dust, cosmic rays). The field structure in the heliosphere is complex and not fully understood. Direct measurements of the pristine ISM magnetic field would be critical to understanding our heliosphere. The critical measurements are the magnetic field orientation, strength, and its variability and turbulence.

Together with the magnetic field strength, measurements of the fundamental properties of the gas, such as temperature and turbulent structure, are needed to evaluate the pressure balance in the LISM. Knowledge of how pressure is balanced is necessary to understand the origins and evolution of the ISM. It is a long-standing mystery for all ISM science, and really requires *in situ* measurements to fully characterize the small-scale variations, at the source of the pressure drivers.

These one-of-a-kind measurements provide our best determination yet of the interstellar spectra of cosmic ray ions and electrons. It is important to extend direct measurements of interstellar ions and electrons, especially to lower energies, but also to higher energies. Interstellar measurements of radioactive isotopes in cosmic rays are important to distinguish cosmic ray acceleration and transport models, including measurements of the cosmic ray lifetime in the Galaxy, and of the average density of material in the cosmic ray storage region.

### 3.7 *Science of the Outer Solar System: Zodiacal Background and Interplanetary Dust*

Access to the outer solar system provides unprecedented opportunities to measure properties of the interplanetary dust (IPD) clouds. These clouds are produced by either continuing collisional cascades that convert larger bodies into smaller ones, or by evaporation of cometary material. In either case, dust continually fills the interplanetary space. While smaller particles are ejected from the solar system by solar radiation pressure, larger ones may spiral sunward under the action of the Poynting-Robertson effect. As a result, dust populations are continually replenished by a rumble of activity, present over time scales of millions of years.

The main sources of zodiacal dust in the inner solar system and likely Kuiper belt dust as well are the main asteroid belt, lying between the orbits of Mars and Jupiter, and the Kuiper belt, lying beyond the orbit of Neptune, together with cometary contributions. Dust belts produced by similar processes are also commonly found around nearby stars, tracing planetesimals in these systems just as they do in our solar system. A study of the IPD is of broader scientific interest because of its connection with extrasolar planetary systems; it could tell whether our own planetary system is an outlier in some important way.

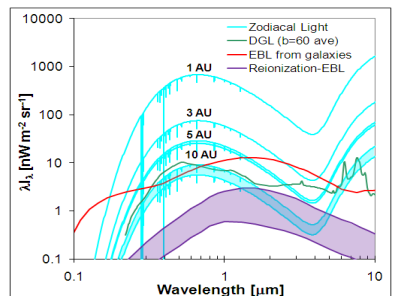

Estimated brightness of the Zodiacal light and other components.

Access to the outer solar system allows for a significant reduction in zodiacal brightness, enabling precise measurements of the Extragalactic Background Light (EBL). It also allows for a deep search for photons from reionization, which is critical to interpret cosmological data. As we travel toward Saturn's orbit and beyond, we can monitor zodiacal light continually, looking for resonant structures and other features that may represent the processes shaping the IPD cloud and thus may be important for exoplanetary systems. As the probe moves outward, sunlight that is scattered and reradiated by planetesimals will drop in intensity, allowing the clearest look ever at the brightness of the extragalactic sky at visible and IR wavelengths.

An instrument flying toward the outer solar system could measure the radial distribution of the IPD and map resonant enhancements and band structures in the zodiacal dust influenced by planetary bodies. It could study the compositional distribution of dust and determine if it arises from comets, asteroids, or both, from the inner to the outer solar system. This instrument would provide important data on (i) how the dust diffuses outward to fill the solar system; (ii) how dust properties (i.e., density, composition, size, etc.) change as a function of radial position; (iii) how the observed dust populations compare with those seen in exoplanetary systems. Such important science investigations may be easily performed even with first-generation sailcraft.

### 3.8 *Cosmic Background & Epoch of Re-ionization*

Zodiacal light is the dominant factor in all Extragalactic Background Light (EBL, the integrated brightness from all photon sources since the Big Bang) measurements obtained at 1 AU. There is a need to perform these measurements from outside the zodiacal dust cloud, beyond the orbit of Saturn. Conducting EBL observations with sufficient resolution to identify individual stars makes it possible to strongly suppress these foregrounds, leading to important results.

From their vantage points deep in the outer solar system and beyond, sailcraft can study fundamental questions in astrophysics and cosmology. By measuring the EBL, they will be able to address questions such as how the Universe originated and evolved to create its galaxies, stars, and planets that we see today. The EBL is a cornerstone measurement needed to probe the fossil record of star formation and galaxy assembly from the first stars to the present day.

We could make the first precise measurements of the EBL at optical and near-infrared wavelengths, especially at those corresponding to the peaks of emitted stellar radiation. The data on the low zodiacal foreground brightness observed from the outer solar system can be processed together with multiband spectral information. Imaging data at arcsecond resolution can be used to remove the galactic and zodiacal foregrounds that have limited previous EBL measurements from Earth's orbit at 1 AU from the Sun (Primack et al. 2009).

### 3.9  *Probing Planet 9*

The clustering of orbits for a group of extreme trans-Neptunian objects suggests the existence of an unseen planet of mass $M \sim 5\text{-}10 M_\oplus$, the so-called *Planet 9*, at a distance of ~400-800 AU from the Sun (Batygin et al. 2019). Direct electromagnetic searches have not yet detected this *Planet 9*.

Witten (2020) proposed to use a relativistic sailcraft, envisioned by the Breakthrough Starshot, to indirectly probe *Planet 9* through its gravitational influence on the probe's trajectory. A similar proposal was made earlier to measure the mass of planets via interferometry by an array of Starshot spacecraft (Christian & Loeb 2017) and also to probe *Planet 9* (Loeb 2019). Lawrence & Rogoszinski (2020) considered the transverse effect of gravity and derived the angular deflection of the probe's trajectory to be $\sim 10^{-9}$ rad. They argued that an angular deflection of this magnitude can be measured with an Earth-based or near-Earth based telescope suggesting that their method is better than measuring the time delay because the transverse effect is permanent, whereas the time delay is only detectable when the spacecraft passes close to *Planet 9*. Other methods to measure properties of *Planet 9*, including its mass and radius, were also considered. For example, Schneider (2017) studied stellar occultations and gravitational microlensing for that purpose.

Both Witten (2020) and Lawrence & Rogoszinski (2020) assumed that the spacecraft is moving on a geodesic trajectory from Earth that is shaped only by gravity. They did not consider the effects of drag or electromagnetic forces from the interaction of the spacecraft with the ISM.

Hoang & Loeb (2020) compared these effects to the gravitational force induced by *Planet 9*. They found that due to interaction with the ISM, a sailcraft would experience fluctuations in the drag and magnetic forces that will dominate over *Planet 9*'s gravitational influence at speeds well above $10^{-3}c = 300$ km/s. Since the drag force depends on the local density, any fluctuations due to ISM turbulence would cause noise in the force well beyond the desired gravitational signal. Collisions with dust grains and fluctuations in the spacecraft charge and interstellar magnetic field will all contribute to an unpredictable and time-dependent noise level that will be difficult to overcome.

These conclusions open an interesting opportunity of using subrelativistic sailcraft to probe these effects and explore the dust and particle background in the ISM. After the acceleration phase, the ''ChipSats'' are coasting, which means that any orbit perturbations must be a result of external forces. By simply tracking the ChipSats, the nature of such forces may be studied. Adding a magnetometer to the sailcraft would allow the study of the space traversed while keeping the data return scalable. Imaging instruments will further enhance the discovery capabilities.

### 3.10  *Studying Other Objects in the Kuiper Belt and Oort Cloud*

The Kuiper belt is a disc-shaped region beyond the orbit of Neptune, extending to 50 AU from the Sun. Since its discovery, the number of known Kuiper Belt Objects (KBOs) much increased. More than $10^5$ KBOs over 100 km in diameter are thought to exist. The recently discovered dwarf planets – Haimea, Makemake, Eris, and Quaoar – all provide interesting targets for exploration. These objects orbit the Sun at the very edges of the solar system at distances ranging 40–90 AU. Another interesting target would be Sedna, or other objects from the new family of the extended scattered disc objects recently discovered with semimajor axes extending well beyond 100 AU.

A mission reaching out to the outer solar system presents a unique opportunity to fly by a large KBO. Various KBO candidates were studied for a near-term mission with MakeMake, Haumea, and Quaoar determined to be of high science value. Out of these three, Quaoar was studied the most. Quaoar is one of the most interesting KBOs. It is in a transition between the large, volatile-dominated, atmosphere-bearing planetesimal and the typical mid-sized volatile-poor object. For most of its history, it had a methane atmosphere, but it is now in the last stages of losing it. Most likely, its surface is patchy in methane frost with the methane being mostly cold-trapped near the poles or in craters. The processes related to atmospheric loss in the outer solar system are poorly known, so Quaoar offers an interesting opportunity to see the process in its late stages. Given its size, Quaoar may have ancient cryovolcanic flows on the surface offering clues on its history.

To investigate these processes, we can make full global imaging in broadband colors, which may be achieved with a swarm of sailcraft. In addition, we may study Quaoar's interior using one of the sailcraft as an impactor. Imaging the crater would be an interesting probe into surface conditions. Plume spectroscopy could explore subsurface composition.

Key science questions will include: (i) What is the fraction of cryovolcanic coverage? (ii) When was the last activity? (iii) What is the depth and coverage of methane? (iv) What is the crater count and ages of frosty surfaces? (v) What is the spatial distribution of volatiles? (vi) What is the mass ratio of Quaoar/Weywot? (vii) Are there additional moons? (viii) What is Quaoar's interior the structure? (ix) What is the structure of Quaoar's atmosphere? Similar questions could also help study Haumea and MakeMake. Swarms of sailcraft may be used to carry out these investigations.

### 3.11 *Probing the Nature of Gravity on the Way Out from the Solar System*

Sailcraft on fast hyperbolic trajectories may be used to test the foundations of relativistic gravitation on scales never before attempted. Precision navigation of spacecraft from 1 AU to 100 AU enables a very powerful test of long-distance modifications of gravity. These measurements can help refine constraints on the parameters of a Yukawa-type extension of Newtonian gravity: effects due to such forces increase with distance, leading to a possible improvement by a factor of 100 compared to current results (Turyshev 2008). Such an experiment can also help improve limits on the post-Newtonian parameters of generalized theories of gravitation, in particular Eddington's $\gamma$-parameter, which measures the effects of spatial curvature. Additionally, limits on the minimum acceleration $a_0$ and constraints on the interpolating function $\mu(g/a_0)$ of Modified Newtonian Dynamics (MOND) (Blanchet & Novak, 2011; Milgrom, 2014) can be refined, along with possible violations of the weak equivalence principle.

Such an experiment will allow placing stringent limits on any possible violations of general theory of relativity while probing for the presence of dark matter in the outer solar system. It will be important for validation of various theories of modified or massive gravity that provide alternatives to dark matter or a cosmological constant. Alternatively, the results could lead to spectacular new discoveries. Experimental confirmation of new fundamental forces would provide an important insight into the physics beyond the Standard Model. These results will bridge the presently extant gap between the data in the solar system and astronomical data obtained on scales that are orders of magnitude larger, involving the dynamical gravitational behavior of star clusters and galaxies.

### 4. Conclusions

Recently, there has been a significant maturation in the technology of smallsats and onboard sensors. Ongoing miniaturization of electronic and mechanical components brought unprecedented reductions in spacecraft size and mass and resulted in successful applications of smallsats in interplanetary experiments. Clearly, much more is possible! Due to their small form factor, smallsats introduce a new paradigm in spacecraft design, relying on agile development, rapid iterations and massive redundancy, which leads to major cost reduction and improved risk posture. In addition, solar sail propulsion demonstrated very impressive results in space, paving the way for laser-pushed sailcraft to enter the scene of planetary exploration.

This progress opens an opportunity for transformational advancements in the space sciences using sailcraft on fast, highly energetic trajectories. This vision may be achieved with standalone science missions or through partnerships with industry and philanthropic foundations that make use of the increasing number of small satellites. The scientific community would benefit tremendously from the data acquired by these novel miniaturized intelligent space systems that offer the capability of rapid access to the outer solar system via light-pushed spacecraft.

Miniaturized intelligent space systems offer major advantages for science investigations. We expect that objects of major scientific significance, like debris from extrasolar planets and pristine building blocks of our solar system, as well as the possibility to image exoplanets in our interstellar neighborhood will foster novel approaches to space exploration. Private companies with internal and government support are paving the way for large-scale manufacturing of capable space platforms at low recurring costs and offer a business model for future endeavors. With proper coordination among key players, science could benefit from these developments in the coming decade, providing a bold vision for renewed planetary exploration.

**Acknowledgements.** This work was supported in part by grants from the Breakthrough Prize Foundation and the National Aeronautics and Space Administration.

*References:*

Batygin, K., Adams, F. C., Brown, M. E., and Becker, J. C. (2019), PhR, 805, 1

Blanchet, L., and Novak, J., (2011), Mon. Not. R. Astron. Soc. 412, 2530–2542

Christian, P., and Loeb, A. (2017), ApJL, 834, L20

Hoang, T., and Loeb, A., (2020); arXiv:2005.01120

Huybrighs, H. L. F., et al., (2020), Geophys. Res. Lett., https://doi.org/10.1029/2020GL087806

Jewitt, D., et al., (2017), ApJ Lett., 850, L36

Labeyrie, A., 1996, A&A Sup. Ser, 118, 517

Labeyrie, A., 1999, Science, 285, 1864

Lawrence, S., and Rogoszinski, Z., (2020); arxiv:2004.14980

Lingam, M., and Loeb, A., (2019), Acta Astronautica, 168, 146; arXiv:1911.02765

Loeb, A. (2019), SciTech Europa Quarterly, 31, 1; https://www.cfa.harvard.edu/~loeb/Loeb_Starshot.pdf

Lubin, P., and Hettel, W., (2020), Acta Futura 12, 9-45

Marcos, C., et al., (2018), MNRAS Lett., 476, L1-L5; arXiv:1802.00778

Millan, R. M., et al., (2019), Adv. in Space Res. 64, 1466–1517

Milgrom, M., (2014), Scholarpedia, 9(6):31410

Xplore (2020): https://www.xplore.com/press/releases/2020/04.28.2020_Xplore_SGLF_NASA_NIAC_Phase_III.html

Opher, M., Loeb, A., et al., (2020), Nat. Astron., (2020), https://doi.org/10.1038/s41550-020-1036-0

Parkin, K., (2018), Acta Astronautica, 152, 379; arXiv:1805.01306

Primack, J. R., et al. (2009), AIP Conf. Proc. 1085 1, 71-82; arXiv:0811.3230

Siraj, A., Loeb, A., (2019), ApJL, doi: 10.3847/2041-8213/ab042a; arXiv:1811.09632v5

Schneider, J., (2017), PASP, 129, 980

Sparks, W. B., et al., (2016), ApJ, 829(2):121

Stone, E. C., et al., (2015), https://www.kiss.caltech.edu/final_reports/ISM_final_report.pdf

Turyshev, S. G., (2008), Annu. Rev. Nucl. Part. Sci. 58, 207-248; arXiv:0806.1731

Turyshev, S. G., and Toth, V.T., (2017), Phys. Rev. D 96, 024008; arXiv:1704.06824

Turyshev, S. G., and Toth, V.T., (2020); arXiv:2002.06492

Turyshev, S. G., et al., (2020); arXiv:2002.11871

Witten, E., (2020); arXiv:2004.14192

# Appendix: *Fiducial Parameters of a Laser-pushed Light Sail System to $10^{-3}c$*

A $10^{-3}c$ (63 AU/yr) cruise velocity enables a sailcraft to reach Pluto within a year, the heliosphere in 2 years, and interstellar space in 5 years. Below we provide some fiducial parameters for a laser-pushed sailcraft mission (Parkin 2018). The inputs to the precursor mission point designs are summarized in Table 1. The 1.06 μm wavelength is consistent with ytterbium-doped fiber amplifiers. An initial sailcraft displacement of 300 km is consistent with a low Earth orbit from which the sail craft is entrained by a low-power beam.

Table 1: Model inputs for $10^{-3}c$ precursor

| |
|---|
| 0.001 c target speed |
| 1.06 μm wavelength |
| 300 km initial sail displacement from laser source |
| 1, 10, and 100 kg payload |
| 0.2 g/m² areal density |
| $10^{-5}$ spectral normal absorptance at 1.06 μm |
| 70% spectral normal reflectance at 1.06 μm |
| 625 K maximum temperature |
| 0.01 total hemispherical emittance (2-sided, 625 K) |
| 0.1 $/W laser cost |
| $10k/m² optics cost |
| $0.04/kWh grid energy cost |
| $4.0/kWh storage cost |
| 50% wallplug to laser efficiency |
| 70% of beam power emerging from top of atmosphere |

Sail mass is calculated by the system model (Parkin 2018) based on the value of sail size chosen by the optimizer combined with the areal density given as an input. Similar to the $0.2c$ design, this design assumes a photonic crystal sail material with the same thermal, mass, optical properties, except for a less ambitious absorptance of $10^{-5}$. Absorptance and reflectance remain constant throughout trajectory integration.

Costs must be reduced from present values for a $0.2c$ mission to be feasible. The cost factors in Table 1 are thresholds and not predictions; the thresholds can only be met with trial-and-error over the coming decade of research. The thresholds can be met in principle: Laser cost is 0.1 $/W, factors of ~$10^2$–$10^3$ lower than the present cost, yet a factor of 10 greater than microwave oven magnetrons, whose production has been nearly completely automated, and well above raw material costs. Optics cost is $10k/m², two orders of magnitude lower than the present cost for diffraction-limited optics, but consistent with significant production line automation. This estimate is comparable with the cost of radio telescope apertures, and again well above raw material costs.

The $4/kWh energy storage cost is consistent with grid-supplied power (as opposed to stored) with the beamer economically optimized for 100 missions (100 times the $0.04/kWh grid energy cost). This is an important choice, as it raises the beamer capex so that overall cost is minimized. Upon running the system model using the inputs in Table 1, the optimizers converge to the values given in Table 2.

Table 2: Model outputs for $10^{-3}c$ precursor

| Payload Mass [kg] | 1 | 10 | 100 |
|---|---|---|---|
| Beamer capex [$M] | 450 | 1,300 | 4,300 |
| of which lasers [$M] | 300 | 890 | 2,900 |
| of which optics [$M] | 150 | 450 | 1,400 |
| Sail diameter [m] | 45 | 97 | 200 |
| Sail+payload mass [kg] | 1.3 | 11 | 110 |
| Peak irradiance [kW/m²] | 1,400 | 840 | 670 |
| Beamer diameter [m] | 140 | 240 | 430 |
| Peak radiated power [GW] | 3 | 8.9 | 29 |
| Peak exitance [kW/m²] | 200 | 200 | 200 |
| Beam duration [d] | 0.57 | 1.5 | 4 |
| Pulse energy cost [$M] | 3.3 | 26 | 220 |
| Pulse energy [GWh] | 83 | 640 | 5,600 |
| Syst. energy efficiency [%] | 0.02 | 0.022 | 0.024 |
| Peak pressure [mPa] | 6.3 | 3.9 | 3.1 |
| Peak force [N] | 9.9 | 29 | 94 |
| Peak acceleration [m/s²] | 7.6 | 2.5 | 0.89 |
| Final acceleration [m/s²] | 2 | 1.3 | 0.67 |
| Final distance [AU] | 0.056 | 0.14 | 0.35 |

If the payload is a lumped mass, not integrated uniformly into the sail, then structural members to mediate forces between the payload and the sail will need to be designed, and their masses deducted from the payload mass. For the 100 kg payload, 29 GW is likely more than a national grid can provide over a 4-day period. At this time, commercial ultra-high voltage direct current transmission lines carry up to 12 GW and are installed along select geographical paths.

If a sailcraft is to be accelerated out of the ecliptic plane in the direction of the Earth's North or South poles, and if some cloud cover can be tolerated, then a single beamer can be used for the entire time. Otherwise, to maintain a beam for more than a few hours, several beam directors are needed to hand off from one to the next as the Earth rotates or there is cloud cover. It may be preferable to use a single beam director and accelerate the sailcraft for only a portion of each day. Based on the data in Table 2, the energy needed to accelerate a sailcraft to $10^{-3}c$ costs 2−3M$ per kilogram of payload for payload sizes in the range of 1−100 kg.